\documentstyle[epsf,rotate]{mn}

\title[The planets capture model of V838 Mon]
{The planets capture model of V838~Monocerotis: conclusions for
the penetration depth of the planet/s}

\author[A. Retter et al.]
{Alon Retter$^{1}$\thanks{Email: retter@astro.psu.edu; 
bzhang@physics.unlv.edu; siess@astro.ulb.ac.be; levinson@wise.tau.ac.il},
Bing Zhang$^{2}$, Lionel Siess$^{3}$ and Amir Levinson$^{4}$\\
$^{1}$Department of Astronomy and Astrophysics, Penn State University, 
525 Davey Lab, University Park, PA, 16802-6305, USA\\
$^{2}$Department of Physics, University of Nevada, Las Vegas, 4505 South 
Maryland Parkway, Las Vegas, NV 89154-4002, USA\\
$^{3}$Institut d'Astronomie et d'Astrophysique, Universit\'e 
Libre de Bruxelles, CP 226, 1050 Brussels, Belgium\\
$^{4}$School of Physics and Astronomy, Tel Aviv University, Tel Aviv 69978, Israel}

\date{Accepted ???. Received ???; in original form ???}

\begin{document}

\maketitle

\begin{abstract}

V838~Mon is the prototype of a new class of objects. Understanding the 
nature of its multi-stage outburst and similar systems is challenging. 
So far, several scenarios have been invoked to explain this group of 
stars. In this work, the planets-swallowing model for V838~Mon is 
further investigated, taking into account the findings that the 
progenitor is most likely a massive B type star. We find that the 
super-Eddington luminosity during the eruption can explain the fast 
rising times of the three peaks in the optical light curve. We used 
two different methods to estimate the location where the planets were 
consumed. There is a nice agreement between the values obtained from 
the luminosities of the peaks and from their rising time scale. We
estimate that the planets were stopped at a typical distance of one 
solar radius from the center of the host giant star. The planets-devouring 
model seems to give a satisfying explanation to the differences in the 
luminosities and rising times of the three peaks in the optical light 
curve of V838~Mon. The peaks may be explained by the consumption of 
three planets or alternatively by three steps in the terminal falling 
process of a single planet. We argue that only the binary merger and 
the planets-swallowing models are consistent with the observations of 
the new type of stars defined by V838~Mon.


\end{abstract}



\section{Introduction}


V838~Mon had an extraordinary multi-stage outburst during the beginning 
of 2002. Fig.~1 taken from Retter \& Marom (2003) displays its 
optical light curve zoomed on the three months of the eruption. Imaging 
revealed the presence of a spectacular light echo around this object 
(Bond et al. 2003). The amplitude of the outburst in the optical band 
was about 9.5 mag -- at the low range of nova outbursts. Novae are 
thermo-nuclear-runaway events in which a white dwarf ejects its outer 
shell, which was accreted from its main-sequence secondary star over 
several thousand years. The post-outburst spectroscopic observations 
of V838~Mon showed, however, that it was very red throughout the 
eruption and long after it ended (Munari et al. 2002; Banerjee \& 
Ashok 2002; Kimeswenger et al. 2002; Evans et al. 2003; Kaminsky \& 
Pavlenko 2005; Tylenda 2005). This is inconsistent with an exposed 
hot white dwarf in novae. 

\begin{figure}

\centerline{\epsfxsize=3.3in\epsfbox{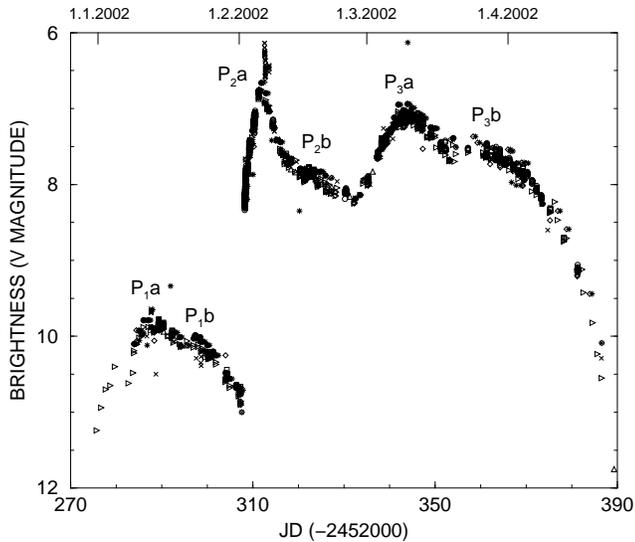}}

\caption{The visual light curve of V838~Mon during the first three 
months after discovery, taken from Retter \& Marom (2003). The object 
showed three peaks separated by about a month. Retter \& Marom (2003) 
argued that the three local maxima (P$_i$a, i=1,2,3) are accompanied 
by secondary shallower spikes (P$_i$b, i=1,2,3), supporting the idea 
that the star experienced three similar events, which presumably are 
the swallowing of three planets. An alternative possibility, which
we propose in this work, is that the three peaks represent three 
steps in the falling process of a single planet at different radii.} 

\end{figure}

Initial estimates of the distance to V838~Mon were below 1 kpc. In 
recent papers there is, however, a general consensus that it is in 
the range $6-10$ kpc (Bond et al. 2003; Tylenda 2004; Lynch et al. 
2004; Van-Loon et al. 2004; Crause et al. 2005; Munari et al. 2005; 
Deguchi, Matsunaga \& Fukushi 2005).
Evans et al. (2003) and Retter \& Marom (2003) concluded that 
the progenitor star of V838~Mon probably had a radius of $\sim 8 
R_\odot$, a temperature of $\sim 7,300$ K and a luminosity of $\sim 
100-160 L_\odot$. Tylenda, Soker \& Szszerba (2005) presented a 
detailed analysis of the progenitor. They argued that V838~Mon is 
likely a young binary system that consists of two $5-10 M_\odot$ B 
stars and that the erupting component is a main-sequence or pre-main 
sequence star. They also estimated for the progenitor a temperature 
of $\sim 4,700-30,000$ K and a luminosity of $\sim 550-5,000 L_\odot$. 
Tylenda (2005) adopted a mass of $\sim 8 M_\odot$ and a radius of 
$\sim 5 R_\odot$ for the progenitor of V838~Mon. There is additional 
supporting evidence that the erupting star belongs to a binary system 
with a hot B secondary star (Munari \& Desidera 2002; Wagner \& 
Starrfield 2002; Munari et al. 2005).

Spectral fitting suggested that V838~Mon had a significant expansion 
from a few hundreds to several thousands stellar radii in a couple of 
months during the outburst (Soker \& Tylenda 2003; Retter \& Marom 
2003; Tylenda 2005; Rushton et al. 2005b). Interferometric observations 
at the end of 2004 with the Palomar Testbed Interferometer confirmed 
the huge radius of the post-outburst star with an estimate of 
$1,570 \pm 400 R_\odot$ (Lane et al. 2005).

Rushton et al. (2003) set an upper limit of $0.01 M_\odot$ for 
the ejecta from the absence of molecular emission from V838~Mon.
Using infrared observations and assuming a model of a spherically 
symmetric shell, Lynch et al. (2004) estimated, however, that the 
mass ejected in the outburst of V838~Mon is about $0.04 M_\odot$. 
Tylenda (2005) concluded that the total mass lost by V838~Mon is 
$\sim 0.001 - 0.6 M_\odot$ and Tylenda \& Soker (2006) adopted a 
range of $0.01-0.1 M_\odot$. The high infrared excess indicates 
multiple episodes of ejection of large amounts of material during 
the outburst of V838~Mon (e.g., Crause et al. 2003). It seems that 
the mass of the matter ejected during the eruption event is well 
above the typical values in nova outbursts, which are about 
10$^{-5} - 10^{-4} M_\odot$ (e.g., Warner 1995). 

The estimates of the expansion velocities of the ejecta of about 50 
-- 500 km s$^{-1}$ (Osiwala et al. 2003; Crause et al. 2003; Kipper et 
al. 2004; Rushton et al. 2005a) are at the low range of nova outbursts. 
Rushton et al. (2005a) inferred from infrared observations that some 
material began falling back into the star in 2003 December and Tylenda 
(2005) described the decline of V838~Mon by a collapsing envelope 
of $\sim 0.2 M_\odot$. Banerjee et al. (2005) found water lines in 
near-infrared spectra of V838~Mon, and related them with a region
around the star, with a temperature of $\sim 750-900$ K, which appears
to be cooling in time. SiO maser emission from V838~Mon was detected by 
Rushton et al. (2005a), Deguchi et al. (2005) and Claussen et al. (2005).


Van-Loon et al. (2004) announced a weak detection of multiple shells 
around V838~Mon using archival IRAS and MSX infrared data. They thus 
proposed that it is a low-mass asymptotic giant branch (AGB) star 
that had several thermal pulses in the past. This result is, however, 
inconsistent with the presence of a young B companion star. Tylenda 
(2004) investigated the structure of the dust distribution in the 
vicinity of V838~Mon. Near the central object he detected a strongly 
asymmetric dust-free region, which he interpreted as produced by a 
fast wind from the central system. Tylenda (2004) proposed that the 
asymmetry implies that V838~Mon is moving relative to the dusty 
medium, and concluded that the dust illuminated by the light echo is 
of interstellar origin rather than produced by mass loss from V838~Mon 
in the past. Crause et al. (2005) stated that the dust is likely in 
the form of a thin sheet distant from the star, and thus supported the 
idea that this material is interstellar. Tylenda et al. (2005b) criticized 
and questioned Van-Loon et al.'s (2004) results and further argued that 
V838~Mon is made of a binary systems with two hot stars, and that the 
progenitor cannot be a red star. 

To summarize, V838~Mon had a spectacular outburst, which has attracted 
many researchers (both observers and theoreticians), but several 
features of this unique object are still controversial and somewhat 
confusing.

\subsection{Models for the outburst}

Soon after its outburst, V838~Mon was recognized as the prototype 
of a new class of stars (Munari et al. 2002; Bond et al. 2003), 
which currently consists of three objects: M31RV (Red Variable in M31 
in 1988; Rich et al. 1989; Mould et al. 1990; Bryan \& Royer 1992), 
V4332~Sgr (Luminous Variable in Sgr, 1994; Martini et al. 1999), and 
V838~Mon (Peculiar Red Variable in 2002), plus three candidates -- 
CK~Vul, which was identified with an object that had a nova-like event 
in the year 1670 (Shara \& Moffat 1982; Shara, Moffat \& Webbink 1985; 
Kato 2003; Retter \& Marom 2003), V1148 Sgr, which had a nova outburst 
in 1943 and was reported to have a late type spectrum (Mayall 1949; 
Bond \& Siegel 2006), and the peculiar variable in Crux that erupted 
in 2003 (Della Valle et al. 2003). 

So far, seven explanations for the eruption of these objects have been 
supplied. The first invokes a nova outburst from a compact object, 
which is embedded inside a common red giant envelope (Mould et al. 
1990). In the second model, an atypical nova explosion on the surface 
of a cold white dwarf was suggested (Iben \& Tutukov 1992; Boschi \& 
Munari 2004). Soker \& Tylenda (2003) proposed a scenario in which a 
main sequence star merged with a low-mass star. This model was lately 
revised by Tylenda \& Soker (2006). Van-Loon et al. (2004) argued that 
the eruption was a thermal pulse of an AGB star. Munari et al. (2005) 
explained the outburst of V838~Mon by a shell thermonuclear event in 
the outer envelope of an extremely massive (M $\sim 65 M_\odot$) B 
star. Recently, Lawlor (2005) proposed another mechanism for the 
eruption of V838~Mon. He invoked the born-again phenomenon to explain 
the first peak in the light curve and altered the model by adding 
accretion from a secondary main-sequence star in close orbit to 
explain the second peak in the optical light curve of V838~Mon.

A promising model for the peculiar eruption of V838~Mon was suggested 
by Retter \& Marom (2003). They showed that the three peaks in its 
optical light curve have a similar double-shaped structure (see Fig.~1) 
and interpreted them as the devouring of three Jupiter-like massive 
planets by an expanding host star that leaves the main sequence. They 
proposed that it is either a red giant branch (RGB) or an AGB star. 
The planets-swallowing scenario had been analyzed in detail by Siess 
\& Livio (1999a, b), however, their simulations indicate relatively 
long time scales for the process. 

Retter \& Marom (2003) calculated that the gravitational energy 
released by a Jupiter-like planet that reaches a distance of one
solar radius from the center of a solar-like parent star is sufficient 
to explain the observed eruption. In addition, they found that the 
time scales of the outburst of V838~Mon could be explained by 
this process. Retter \& Marom (2003), therefore, argued that the 
planets-devouring model is generally consistent with the observed 
properties of this object, including its possible binary nature 
mentioned above. This is since planets have been observed in binary 
systems (e.g., Marcy et al. 2005; Mugrauer et al. 2005; Schneider 
2006). As discussed above, it was found that the progenitor of 
V838~Mon is very likely a young B star. The planets-swallowing 
scenario is consistent with a B-type progenitor as well. The 
initial slow expansion of the parent star may occur as a result 
of the natural stellar evolution after leaving the main sequence. 


In the following, we adopt the planets-swallowing model for V838~Mon 
and further explore this scenario and its implications. The progenitor 
of V838~Mon is very likely a massive B type star. It should be kept 
in mind, however, that other types of giant stars, namely RGB and AGB 
stars, are very likely applicable to the other members in this group. 

\section{Where are the planets stopped?}

Within the planets-devouring model for V838~Mon, we can estimate the 
distance from the center of the host star where the swallowing process 
takes place. The planet is assumed to be engulfed by the stellar 
envelope. The consumption is defined as the point in space and time 
where and when the impacting planet has come to rest relative to the 
stellar envelope, i.e. when it has transferred all (or most of) its 
kinetic energy to the parent star. 

The motion of a secondary star that orbits inside the envelope of a 
primary giant star was discussed in detail by Livio \& Soker (1984), 
Soker (1998) and Siess \& Livio (1999a, b). It is generally accepted 
that there is a limit for the secondary mass of $\sim 1-10 M_J$, 
where $M_J$ is the Jovian mass, above which the planet can survive and 
reach the stellar core before evaporation. The physics of the spiraling 
process is extremely complex and is outside the scope of this paper. 
Therefore, in this work we simply assume that the planet manages to 
arrive to the stellar core and that it does not dissolve earlier. The 
planetary mass is assumed to be constant, and we ignore a few possible 
effects in the inward-falling process such as evaporation, mass loss, 
mass accretion, stellar expansion, influence of the propagating shocks 
and energy deposition by the planet. 

Let $M_p$ denote the mass of a planet that starts from a large radius 
(say from the stellar edge) and reaches a distance $r_o$ from the 
center of a parent star in a time scale $t$. $M_{in}$ represents the 
stellar mass enclosed within this radius. The luminosity emitted by 
the planet is then approximately:

\begin{equation}
L= \alpha \frac{G M_{in} M_p} {r_o} / t
\end{equation}
where $G$ is the gravitational constant and 0 $\leq \alpha \leq $ 1 
is the energy efficiency. In this equation we neglected the kinetic 
energy of the ejected matter and the gravitational energy of the 
expanded shell or took them into account in the $\alpha$ parameter. 
We note that the estimates of these values (Section 1) are highly 
uncertain. The stopping radius can be expressed by

\begin{equation}
r_o \sim 1 (\frac{\alpha} {0.25}) (\frac{M_{in}} {M_\odot}) 
(\frac{M_p} {M_J}) (\frac{L} {10^5 L_\odot})^{-1} 
(\frac{t} {30 {\, \rm d}})^{-1}{\, } R_\odot           
\end{equation}
where $M_\odot$, $L_\odot$ and $R_\odot$ are the solar mass, luminosity 
and radius respectively.


It is assumed that the energy released during the eruption event 
dominates the process. Using a distance of 8 kpc (Section 1), for 
the first peak in the outburst of V838~Mon $L \sim 3-9 \times 10^4 
L_\odot$ (Retter \& Marom 2003; Tylenda 2005; Rushton et al. 2005b), 
and $t \sim 30$ d (Fig.~1). Assuming $\alpha=0.25$, for a host star 
with a mass of $\sim 8 M_\odot$ (Section 1) we obtain a stopping 
radius of about $8-26 m_p R_\odot$ where $m_p=M_p/M_J$. For the 
second peak, $L \sim 6-13 \times 10^5 L_\odot$ and $t \sim 25$ d, 
so $r_o \sim 0.7-1.6 m_p R_\odot$. For the third peak, $L \sim 3-11 
\times 10^5 L_\odot$ and $t \sim$ 40 d and we find $r_o \sim 0.5-2 
m_p R_\odot$. The gravitational energy of the inner planet at its 
initial radius, which is presumably about $5 R_\odot$ (Section 1), 
cannot be neglected. Taking this effect into account, we find a final
radius of $4-4.5 R_\odot$ for the first peak for $m_p=1$. Note that 
for simplicity we used $M_{in}=8M_\odot$, but since $M_{in}$ is lower 
than the stellar mass, the correct numbers are somewhat smaller. We 
conclude that the planets are probably stopped at a typical distance 
of about one stellar radius from the center of the host star.

\section{The slowing time scale}


The luminosities reached in the three peaks in the optical light 
curve of V838~Mon were $L \sim 0.3-0.9, 6-13$ and $3-11 \times 
10^5 L_\odot$ (Section~2). Thus, the last two peaks were certainly 
brighter than the Eddington luminosity, which is $(4 \pi cGM) / \kappa 
\sim 1 \times 10^5 L_\odot$ for an $8 M_\odot$ star (Section 1), where 
$c$ is the speed of light and $\kappa$, the opacity, is taken to be of 
the order of unity. The first faintest peak may also be super-Eddington 
taking into account all uncertainties. We conclude that at least the 
two brightest peaks in the outburst of V838~Mon were super-Eddington,
which is consistent with the findings of Tylenda (2005). Therefore, 
the radiative pressure was larger than the gravitational force and the 
material was thrown away at a high speed, which is governed by the 
unbalance between the radiative and gravity forces. This implies that
the photons reach the outer envelope of the giant star very fast, and
the long-term diffusion Kelvin-Helmholtz time scale is irrelevant for 
the outburst process. Thus, the slowing time scale of the planets, 
which is the time it takes them to lose most of their orbital energy, 
should power the light curve.


Consider a planet with mass $M_p$ and radius $R_p$ moving at
an orbital radius $r_{o}$ inside a stellar envelope with a local 
stellar density of $\rho_o$. The relative velocity between the planet 
and the envelope is of the order of the Keplerian velocity. Thus, 
for a circular orbit, the planet velocity is given by,

\begin{equation}
v_{p}\sim(\frac{GM_{in}} {r_o} )^{1/2} \sim 
400 (\frac{M_{in}}{M_\odot})^{1/2} 
(\frac{r_o}{R_\odot})^{-1/2} {\, \rm km \, s^{-1}}
\end{equation}
The corresponding orbital period is

\begin{equation}
t_{p}=\frac{2\pi r_o} {v_{p}}\sim 
0.1 (\frac{M_{in}}{M_\odot})^{-1/2} (\frac{r_o}{R_\odot})^{3/2} {\, \rm d}
\end{equation}


In Fig.~2 we plot several profiles of B type stars on the main sequence 
when r $ < 30 R_\odot$, and later when they become red giants i.e. when 
r $> 100 R_\odot$ (top panels), and of RGB (middle panels) and AGB stars 
(bottom panels) at different locations along their giant branch as
characterized by their radii. These models were calculated by the code 
of Siess, Livio \& Lattanzio (2002). We find that for most cases, the 
planet velocity is larger than the sound speed at the relevant part of 
the stellar envelope. In regions where the planet moves supersonically, 
it will drive a shock, which propagates into the envelope and eventually 
slows down the planet. The planet will be significantly slowed down and 
spiral into an inner orbit or stopped when it encounters an envelope 
mass of the order of its own mass. Therefore, the slowing range is 
approximately given by,

\begin{figure*}

\centerline{\epsfxsize=4.5in\epsfbox{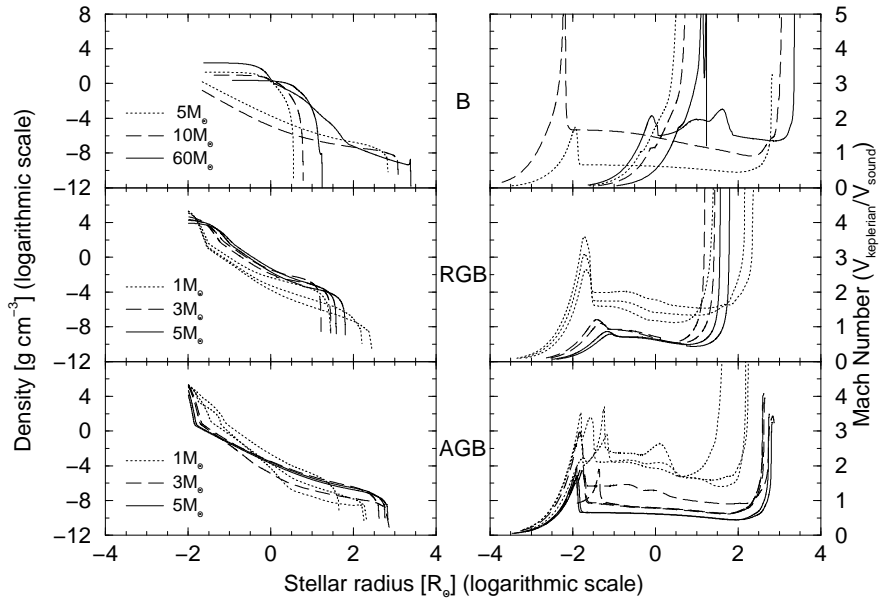}}

\caption{Left-hand graphs: sample density proflies of B, RGB and AGB
stars (top, middle and bottom panels respectively) during different 
stages in their stellar evolution using the code of Siess et al. 
(2002). For B stars plots for 5, 10 and 60$M_\odot$ are marked with 
dotted, long-dashed and solid lines. For RGB and AGB stars, data 
for masses of 1, 3 and 5$M_\odot$ are respectively shown with dotted, 
long-dashed and solid lines. The right-hand panels display the Mach 
number as a function of radius for the same models. The density 
required to stop Jupiter-like planets ($\rho \sim 10^{-3}$ g cm$^{-3}$) 
is obtained at a typical distance of $r \sim 1 R_\odot$ from the center 
of the host star.} 

\end{figure*}

\begin{equation}
l=\frac{M_p} {(\pi R_p^2) \rho_o}=\frac{4}{3}R_p (\frac{\rho_p}{\rho_o})
\sim 10^{10} (\frac{R_p}{R_J}) (\frac{\rho_p}{\rho_o}) {\, \rm cm}
\end{equation}
where $R_J$ represents the Jovian radius and $\rho_p$ the mean planet 
density. The corresponding slowing time is:


\begin{equation}
t_{slow} = \frac{l}{v_p} \sim 
4 (\frac{R_p}{R_J}) (\frac{\rho_p}{\rho_J}) 
(\frac{\rho_o}{10^{-3}})^{-1} (\frac{r_o}{R_\odot})^{1/2} 
(\frac{M_{in}}{M_\odot})^{-1/2} {\, \rm d}
\end{equation}
where $\rho_J$ denotes the Jovian mean density.

The slowing time scale corresponds to the time length when the planet 
loses a significant fraction of the orbital energy. In the light curve, 
this should be expressed by the rise times of the peaks. The rising 
times of the three peaks in the optical light curve of V838~Mon can be 
respectively estimated as about 12, 4 and 10 days (Fig.~1). Thus, a 
local stellar density of the order of $\rho_o \sim 10^{-3}$ g cm$^{-3}$ 
is required at the location where the planets are consumed. From 
Fig.~2 we find that this value is obtained in the range $r \sim 0.1-12 
R_\odot$ for B stars. The upper limit corresponds to a mass of $60 
M_\odot$, which was proposed by Munari et al. (2005), but it is very 
likely unrealistic for V838~Mon. For a B star with a more reasonable 
mass in the range $5-10 M_\odot$ (see Tylenda et al. 2005; Tylenda 
2005), the required density occurs at $r \sim 0.1-4.5 R_\odot$, and if 
V838~Mon is a B star with a mass of $\sim 8 M_\odot$ (see Section~1), 
we find $r \sim 0.2-4 R_\odot$. Note that higher values are obtained
for lower stellar radii, and that the expansion leads to a shift of the 
`critical' density closer to the stellar core. For RGB and AGB stars, 
the required density is respectively found at $r \sim 0.5-10$ and 
$r \sim 0.2-8 R_\odot$. These values are in excellent agreement with 
the conclusion that the planets are swallowed at a characteristic 
distance of one solar radius from the center of the host star using 
the luminosities of the peaks assuming a reasonable energy efficiency 
(Section~2). 

Fig. 2 shows that the stellar expansion, which occurs as a natural step 
in its evolution, leads to a strong change in the density profile. The 
observed expansion during the outburst of V838~Mon probably causes a 
similar structural change in the density profile. For V838~Mon, the 
expansion of the stellar structure induced by the dissipation process 
will shift the critical density to smaller radii inside the star (e.g., 
Siess \& Livio 1999a,b). Therefore, the expansion of the envelope implies 
that a distant planet has to penetrate closer to the stellar core than a 
close planet, in order to hit the critical density and to stop. Thus, 
it makes sense that the inner planet was consumed at a larger distance 
from the center of the host star than the two other planets (Section 2). 


\section{The signature of planets in stellar envelopes}

The rate at which the planet loses energy is given by its kinetic
energy divided by the slowing time:

\begin{equation}
\dot{E} = \frac{0.5 M_p v_p^2}{t_{slow}} \sim 
10 ( \frac {\rho_o} {10^{-6}}) (\frac{R_p}{R_J})^2 
(\frac{M_{in}}{M_\odot})^{3/2} (\frac{r_o}{30R_\odot})^{-3/2} {\, L_\odot}
\end{equation}
This energy deposit rate is adequate to explain the observed luminosity 
in the eruption of V838~Mon even given a moderate radiative efficiency.


Prior to the outburst, if the planet is well inside the envelope
of a host giant star, this luminosity will be smeared out because of 
the long diffusion time scale of the photons. In giant stars, variations 
can be seen at the start of the process, near the stellar radius, but 
then their amplitude would be much smaller. However, even when the 
planet is far away from the core of the parent star (say at $r_o \sim 
30 R_\odot$ and $\rho_o \sim 10^{-6}$ g cm$^{-3}$) about $10 L_\odot$ 
can be produced by the falling process. This would be a fraction of 
the luminosity of the giant star, which is typically a few 
hundreds-thousands solar luminosities, but it is still detectable. The 
deeper the planet is found inside the stellar envelope, the larger 
the amplitude of the variation becomes, however, it is smeared over a 
longer interval of time. Therefore, we expect to observe quasi-periodic 
variations in giant stars only when their planets start the inward 
spiraling process. Based on this idea, Retter (2005, 2006) recently 
proposed to explain the long secondary periods, which are observed 
in the light curves of RGB and AGB stars and whose nature is unknown 
(Wood, Olivier \& Kawaler 2004), by the presence of planets that orbit 
at the outer edges of their host stars.

For the progenitor of V838~Mon, the following parameters were estimated: 
$M \sim 8 M_\odot$, $R \sim 5 R_\odot$ and $L \sim 550-5,000 L_\odot$ 
(Section 1). For these values the expected quasi-periodicity is 0.4 
days (see Eq.~4), and we find that the amplitude of the variations 
before the outburst could have been of the order 
of the stellar luminosity. Goranskij et al. (2004) checked archival 
photographic observations of the progenitor of V838~Mon during 
1928--1994. They could not find any significant variability in 
the $B$ band images. However, their measurements are based on eye 
estimates that are accurate to about 0.2 magnitude, and therefore,
variations with smaller amplitudes or such that are stronger in the 
red, could have been missed. In addition, we comment that the radial 
acceleration is neglected in our calculations, so the fall could be 
quite fast. The density profile of main sequence B stars is very 
steep and there is a sharp rise of about 10 orders in a fraction of 
a stellar radius (Fig. 2, upper left panel). Thus, the falling planet 
penetrates higher densities quite fast, so the swallowing of the inner 
planet could have been very rapid. Based on our estimates, only a 
fall of $\sim 0.5-1 R_\odot$ is necessary to stop a planet that 
orbits at the edge of a massive B star with a radius of $\sim 5 
R_\odot$ (Section~2). On the other hand, the density profiles of 
giant stars is much shallower, so a planet that orbits in their 
outer envelopes slowly falls inwards and can show the quasi-periodic 
variations discussed above. In this context, we note that the steep 
profiles of main sequence B stars imply that the inner planet was 
very likely consumed near the edge of the host star. Therefore, the 
luminosity emitted in the first peak did not have to be super-Eddington 
(see Section 3) for the outburst to be observed.

According to the planets-devouring model of V838~Mon, prior to 
outburst, when the planet is at the edge of the stellar envelope of
its parent star, the giant host star may show some quasi-periodic 
oscillations in its light curve imposed on a gradual increase in 
the luminosity. This is due to energy released by the inward-falling 
planet. Indeed, such a brightening has recently been detected in the 
light curve of the progenitor of V4332 Sgr, one of stars in this new 
group (Kimeswenger 2006). 


\section{The rate of planet-swallowing events}

We can estimate the rate of V838~Mon-like outbursts within the 
planets-capture model for this phenomenon. We assume that this is 
a natural step in the stellar evolution and that no unique trigger
mechanism is required for this process. We first start with 
solar-like stars. The number of stars in the Milky Way is about 
10$^{11}$. The age of a $1 M_\odot$ expanding RGB or an AGB star 
(there is a small difference of $\sim 10^{8}$ years between the two 
phases) is about 1.2 $\times 10^{10}$ years (Sackmann, Boothroyd \& 
Kraemer 1993). Thus we obtain a number to age ratio of about 8 per 
year for these stars. The number of B type stars with masses of 
$\sim 5-10 M_{\odot}$ (see Section 1) in our galaxy can be estimated 
as about 1\% of the whole population from the initial mass function 
(e.g. Lucatello et al. 2005). Their evolution is, however, much 
faster than solar-like stars, and their age on the main sequence is 
estimated as about $2-9 \times 10^{7}$ years (Siess 2006). Therefore, 
about $10-50$ massive stars in the Milky Way leave the main sequence 
every year. 

The estimate of the frequency of V838-like outbursts in our galaxy
should take into account the ratio of stars with Jupiter-like planets
in close orbits. Marcy et al. (2005) concluded that about 12\% of FGK 
stars have Jupiter-like planets. Assuming that about 5\% of all stars 
host planets at the relevant range of masses and separations and devour 
them, we thus expect about 0.4 such events per year in our galaxy for 
solar-like stars and $\sim 0.5-2.5$ outbursts in massive stars. 

Many V838~Mon-like eruptions are probably missed. This effect can 
be accounted for by a comparison with nova outbursts because the 
observational bias for these two types of events is similar. About 
$5-10$ novae are detected in our galaxy each year while estimates for 
the actual occurrence number of these eruptions range between 11 and 
260 (Shafter 1997). Adopting a reasonable value of 50 galactic novae 
per year, we estimate that a single V838~Mon-like event should be 
detected every $\sim 2-10$ years in all stars. These values are in 
agreement with the current three members and one candidate in this 
group that erupted in the past 20 years (Section~1). Note that the 
wealth of poorly studied novae may hide more V838~Mon-like systems. 
The number of galactic novae that are discovered every year is rising 
fast thanks to many new variability surveys. Therefore, we should 
expect an increase in the frequency of the detection of V838~Mon-like 
events as well.

\section{Discussion}


Within the planets-swallowing model for V838~Mon we used two different 
methods to find the location where the planets were swallowed inside 
the envelope of the host star. There is a nice consistency between 
the estimates obtained from the energy balance and from the stellar
density that is required to stop the planets. We concluded that the
planets were consumed at a characteristic distance of about one solar 
radius from the center of their host star (Sections 2 \& 3). This is 
consistent with the fact that a Jupiter-like planet overflows its Roche 
lobe about $2 R_\odot$ away from the core of a host star with a mass 
of $M \sim 8 M_\odot$ (see Eggleton 1983). This is a rough lower limit 
for the final radius because if the planet was not stopped earlier, it 
would then start transferring most of its mass to the host star and 
would quickly dissolve and cease to exist as an independent body. 

The values we derived for the location of the accretion process compare 
very well with the numbers estimated from the Virial temperature by 
Siess \& Livio (1999a). At such a close proximity to the stellar core, 
the temperature of the stellar envelope exceeds 10$^6$ K. Therefore, 
the eruption may be triggered by extra energy received from the nuclear 
burning of deuterium brought by the falling planets. Another option is 
that the outburst occurred once the planet reached the critical stellar 
density, which is required to significantly slow it down. Hitting denser 
material causes higher energy release and increasing radial acceleration 
component. At a density of $\rho \sim 10^{-3}$ gr cm$^{-3}$ and a distance 
of a few solar radii from the stellar core, the opacity, $\kappa$, becomes 
larger than one. Therefore, the trigger for the event could be when the 
luminosity released by the planet is larger than the local Eddington limit. 
Alternatively, the outburst could be triggered by a sudden inward fall of 
the first planet, maybe because of some kind of tidal instability, perhaps 
due to the proximity of the three planets to the parent star and / or to 
each other, or maybe because of eccentric orbits or due to some gravitational 
influence by the secondary star. The consumption of the inner planet and 
the subsequent expansion of the host star led to the engulfment and the 
swallowing of the two other planets. This idea may supply a simple 
solution to the question `how three planets in close orbits around their 
host star can be stable for a long interval of time?', by speculating 
that they were actually unstable. Next, we explore a different idea.

\subsection{One or three planets?}

The planets-devouring scenario for the outburst of V838~Mon can 
explain the rising time scale of the peaks and their strengths. 
However, one difficulty of this model is that Eq.~(6) probably 
implies that the three planets should be relatively close to each 
other in order to fall within a couple of months as observed; 
unless the swallowing of the inner planets, the subsequent mass 
ejection and stellar expansion, or eccentric orbits somehow boosted 
this process. One would expect that the time between episodes of 
falling planets may be of the order of several years, thousands of 
years, or even more. We comment though that the radial acceleration 
is neglected in our simple approach, and that hydrodynamic simulations 
are required to better describe the falling process of the planets.

An alternative scenario to the multiple planets model is that all three 
peaks were produced by the same planet at different radii, which seems 
to be consistent with our estimates (Sections 2 \& 3). The captured 
planet reaches some critical density in the envelope of the host star 
and triggers an initial event that gives rise to super-radiant bubbles, 
perhaps driven by some convective instability that quickly propagates 
out and then cools radiatively over rather short time scales. As a result 
of the large energy release of the first event, the star expands and the 
density of the material surrounding the planet falls below the limit. 
The on-going fall of the planet brings it again into higher densities and 
another super-Eddington event is triggered. The resulting adjustment of 
the envelope is driven the planet out of the critical density value again, 
etc. The planet finally reaches the nuclear burning shell and dissolves 
or just evaporates as argued in Livio \& Soker (1984). By this process, 
the time scales for the duration of each event and the time intervals 
between events are naturally similar. Note that in this case, the observed
rise time of the peaks, which were discussed in Section~3, represent the
slowing down process of a single planet in three different orbits rather 
than the final stopping times of three planets.


\subsection{A Comparison between the different models}

So far seven models have been suggested for the new phenomenon, which is 
defined by V838~Mon (Section 1.1). Five of them can be easily rejected. 
This is because they are restricted to a single type of stars, while the 
members in the V838~Mon class are of different kinds. It was concluded 
that the progenitor of V838~Mon is a B supergiant while that of V4332~Sgr, 
and most likely M31RV, are red giants (Tylenda et al. 2005a; Bond \& 
Siegel 2006). Therefore, only scenarios that use different types of 
stars could be invoked to explain their eruptions. This includes the 
binary merger and the planet/s-swallowing models.

Recently, Tylenda \& Soker (2006) presented an analysis of the 
observational properties of the V838~Mon-like objects and compared in 
detail the various explanations of their eruptions. They argued that 
thermonuclear models (novae and Helium shell flashes including born 
again AGB) cannot be applied to the outbursts of these objects. The 
major arguments against these scenarios were that the observed spectral 
evolution of these stars, the proposed B type progenitor of V838-Mon 
and the presence of circumstellar non-ionized matter are inconsistent 
with these models. In addition, Tylenda \& Soker (2006) pointed out 
that there is no indication of matter processed by nuclear burning in 
these objects. That work can thus be used as a further argument against 
these five models.

Tylenda \& Soker (2006) argued that only a merger model fits all
observed features of the V838-Mon-like stars. They discussed the 
energies obtained in the outbursts of the three objects in this group. 
For M31RV only an upper limit on the brightness of the progenitor is 
available, so no conclusion can be made. For V4332 Sgr, Tylenda et 
al. (2005a) and Tylenda \& Soker (2006) calculated that its outburst 
can be explained by a merger of a solar-like star and a planet. For 
V838~Mon, however, Tylenda \& Soker claimed that the energy released 
by a planet that falls onto a massive star is not sufficient to 
explain the observed eruption, and thus they invoked instead a merger 
with a low mass ($M \sim 0.10-0.33 M_\odot$) star. They also suggested 
that the outburst may occur as a result of a binary merger with a 
single companion or with two stars. This implies that V838~Mon is a 
member in a multiple system, because it was found that it also has a 
massive B companion (Section 1). Tylenda \& Soker (2006) stated that 
a merger with several physical bodies can better explain the huge 
inflation of the stellar radius.

Tylenda \& Soker (2006) listed three arguments against the 
planets-swallowing scenario of V838~Mon. Two of them are that the 
progenitor is not a red giant and that the two outbursts in February 
-- March might have occurred at the base of the inflated envelope 
within $1-2$ days. Our model is certainly consistent with a B type star 
(Section~1.1). Tylenda (2005) found three different slopes in the radius 
expansion of V838~Mon. Extrapolating these lines to earlier times, he 
suggested that they occur within about 2.5 days. Therefore, he argued 
that the eruption phase in February -- March 2002 originated from a 
single outburst event during the last days of January 2002. This would 
be consistent with our suggestion to explain the outburst by a single 
planet. Tylenda added that this would mean that the expansion velocity 
during the first event (the second peak in the light curve) should be 
about 800 km s$^{-1}$, and twice larger than that in the second event 
(the third peak). However, this seems inconsistent with the observations, 
which indicate that the expansion velocities in the two events are similar 
(Rushton et al. 2005b; Tylenda et al. 2005b). An alternative simple 
interpretation to the different rates in the radius derivative is that 
the expansion of the stellar envelope was slower with larger radii.


The major difficulty to the planets-capture model of V838~Mon raised 
by Tylenda \& Soker (2006) is probably the question whether the 
energy produced by this process is sufficient to explain the observed 
outburst. Tylenda \& Soker argued that the total energy involved in the 
event is about $3-10 \times 10 ^{47}$ ergs. Thus, they concluded that 
the $\sim 8 M_\odot$ star could have merged with a companion star 
with a mass of $\sim 0.10-0.33 M_\odot$. This is about $3-10$ times 
larger than the total mass of three massive Jupiter-like planets. 
Tylenda \& Soker (2006) assumed that the falling planet reaches a 
final distance of about $5 R_\odot$ from the center of its $\sim 8 
M_\odot$ host star. However, we estimate that the consumption occurs 
much deeper, say at a radius of $\sim 1 R_\odot$. Thus, the energy 
released by the planet could easily be about five times larger than 
the estimates of Tylenda \& Soker and even higher if the planet gets 
closer to the core of its host star, if it accretes some matter during 
the fall, or if the stellar mass is larger. Therefore, it seems that 
the energy release by the swallowed planets can account for the 
observed outburst of V838~Mon.


The conclusions of Tylenda \& Soker (2006) are based on very rough 
estimates for the mass of the ejecta and the inflated envelope. 
Tylenda \& Soker respectively used $0.01-0.1$ and $\sim 0.2 M_\odot$ 
for these two quantities. However, these values have large uncertainties. 
The measurements of the ejecta mass span a large range of values 
(Section 1), so the mass of the ejected material and the energy release
could be much smaller than the estimates of Tylenda \& Soker (2006). 
Tylenda (2005) argued that the decline of V838~Mon can be described 
by a collapsing envelope of $\sim 0.2 M_\odot$. Such a mass is larger 
than the total mass of three massive planets, and thus challenges our 
model. However, Tylenda did not give any error for this value. From his 
equation A.20 we conclude that the envelope mass is correlated with the 
third power of the stellar radius ($R^3$). Therefore, a possible error 
of a factor of two in the radius could lead to an envelope mass of $\sim 
8$ times smaller. Taking into account the uncertainties in the other 
parameters, and possible asymmetry, which is very likely to occur in 
the planets-devouring model, and could further complicate the picture, 
the shell mass could easily be about 20 times lower than $0.2 M_\odot$, 
which is consistent with our model even for a single planet. 



The scenarios proposed to explain the outbursts in V838~Mon-like 
stars seem to converge to an explanation by accretion of a second 
mass. The difference between the merger model (Soker \& Tylenda 2003; 
Tylenda \& Soker 2006) and our model (Retter \& Marom 2003; this work) 
stems from the mass of the companion. In the binary merger model, a 
low-mass star merged with the primary star, while in our model $1-3$ 
massive planets were consumed.
It is interesting to note that the two models seem to merge. In 
Section~6.1 we proposed that instead of three planets, three steps 
in the falling process of a single planet may be invoked to explain 
the multi-stage optical light curve of V838~Mon. Tylenda \& Soker 
(2006) suggested that maybe two low-mass stars were merged with the 
primary B star in V838~Mon. In this case, it may have been a rare 
quadrapole system. They also stated that the outburst of V4332~Sgr 
could be explained by a merger of a solar-like star and a planet. 
Therefore, the differences between the two models are relatively 
small.

\section{Summary and Conclusions}

Within the planets-capture model for the eruption of V838~Mon, 
this work yields a `semi-empirical' answer to a very interesting 
question: "how deep into giant stars do planets reach before 
consumption?". The estimates obtained from the luminosities of the 
peaks in the optical light curve of V838~Mon, from the rising time 
scale of a few days of these peaks and from the Roche lobe geometry, 
are in agreement with each other and suggest a final typical stopping 
distance of about one solar radius from the center of the host star. 
This consistency adds further support to the planets-devouring model 
of V838~Mon.


The planets-capture scenario for the eruption of V838~Mon supplies 
a consistent description to the differences in the luminosities 
and rising times of the three peaks in its optical light curve. In 
addition, the engulfment of the nearby planets in the stellar envelope 
and their swallowing is a natural result of the secular evolution of 
solar-like stars and their huge expansion after they leave the 
main-sequence stage. We thus believe that this is a plausible model 
for the spectacular multi-stage outburst of this peculiar object and 
the other stars in its class. We think that only two models among
the many offered so far for the V838~Mon phenomenon are consistent 
with the observations. These are the binary merger scenario and the 
planet/s-swallowing model. These ideas are very similar because both 
invoke the accretion of a secondary mass as an explanation for the 
eruption. Two significant differences between the models are the
energetics involved and the evolutionary status of the donor. The 
issue of energetics may be answered in the future with better 
modelling and / or observations.




\section*{Acknowledgments}

We thank Noam Soker, Mercedes T. Richards, Tim Bedding, J\'{a}n Budaj, 
Sara Maddison and Ariel Marom for several useful discussions and 
comments. Two anonymous referee are acknowledged for many wise remarks 
that significantly improved the paper. This work was partially supported 
by a research associate fellowship from Penn State University. LS is a 
FNRS Research Associate


\begin{thebibliography}{99}

\bibitem{b2} Banerjee D.P.K., Ashok N.M., 2002, A\&A, 395, 161

\bibitem{b2}
Banerjee D.P.K., Barber R.J., Ashok N.M., Tennyson J., 2005, ApJ, 627, L141

\bibitem{b2} Bond H.E., Siegel M.H., 2006, AJ, 131, 984 

\bibitem{b2} Bond H.E. et al., 2003, Nat, 422, 405
%

\bibitem{b3_h} Boschi F., Munari U., 2004, A\&A, 418, 869

\bibitem{b9} Bryan J., Royer R.E., 1992, PASP, 104, 179

\bibitem{b2}
Claussen M., Healy K., Starrfield S., Bond, H.E., 2005, IAUC, 8602, 1


\bibitem{b2} Crause L.A., Lawson W.A., Kilkenny D., van Wyk F., 
Marang F., Jones A.F., 2003, MNRAS, 341, 785

\bibitem{b2} Crause L.A., Lawson W.A., Menzies J.W., Marang F., 2005, 
MNRAS, 358, 1352 


\bibitem{b2}
Deguchi S., Matsunaga N., Fukushi H., 2005, PASJ, 57, L25 

\bibitem{b2} Della Valle M., Hutsemekers D., Saviane I., Wenderoth E., 
2003, IAUC 8185



\bibitem{b9} Eggleton P.P., 1983, ApJ, 268, 368

\bibitem{b9} Evans A. et al., 2003, MNRAS, 343, 1054


\bibitem{b13} 
Goranskij V.P., Shugarov S.Y., Barsukova E.A., Kroll P., 2004, 
IBVS, 5511, 1



\bibitem{b4_b} Iben I.J., Tutukov A.V., 1992, ApJ, 389, 369



\bibitem{b4_d} Kaminsky B.M., Pavlenko Y.V., 2005, MNRAS, 357, 38

\bibitem{b3_b} Kato T., 2003, A\&A, 399, 695

\bibitem{b3_b} Kimeswenger S., 2006, AN, 327, 44

\bibitem{b12} Kimeswenger S., Lederle C., Schmeja S., Armsdorfer B., 
2002, MNRAS, 336, L43

\bibitem{b12} Kipper T., Klochkova V.G., Annuk K., Hirv A., Kolka I., 
Leedj\"{a}rv L., Puss A., Skoda P., Slechta M., 2004, A\&A, 416, 1107

\bibitem{b12} 
Lane B., Retter A., Thompson B., Eisner J., 2005, ApJ, 622, L137

\bibitem{b12} 
Lawlor T.M., 2005, MNRAS, 361, 695


\bibitem{b3_c} Livio M., Soker N., 1984, MNRAS, 208, 763


\bibitem{b3_e}
Lucatello S., Gratton R.G., Beers T.C., Carretta E., 2005, ApJ, 625, 833

\bibitem{b3_e} Lynch D.K. et al., 2004, ApJ, 607, 460

\bibitem[{}]{519}
{Marcy} G.W., Butler R.P., Fischer D.A., Vogt S.S., Wright J.T.,
Tinney C.G., Jones R.A., 2005, PThPS, 158, 24

\bibitem{b3_f} Martini P., Wagner R.M., Tomaney A., Rich R.M., 
della Valle M., Hauschildt P.H., 1999, AJ, 118, 1034

\bibitem{b3} Mayall M.W., 1949, AJ, 54, 191

\bibitem{b3_g} Mould J. et al., 1990, ApJ, 353, L35

\bibitem{b11}
Mugrauer M., Neuh\"{a}user R., Seifahrt A., Mazeh T., Guenther E.,
2005, A\&A, 440, 1051

\bibitem{b3_h} Munari U., Desidera S., 2002, IAUC, 8005

\bibitem{b3_i} Munari U. et al., 2002, A\&A, 389, L51

\bibitem{b3_i} Munari U. et al., 2005, A\&A, 434, 1107


\bibitem{b3_k} Osiwala J.P. et al., 2003, in `Symbiotic stars probing 
stellar evolution', R.L.M. Corradi, J. Mikolajewska, T.J. Mahoney eds., 
(ASP Conference Series, San Francisco, 2002), 240


\bibitem{b4_f}
Retter A., 2005, BAAS, 37, 1487

\bibitem{b4_f} Retter A., 2006, ApJ, submitted

\bibitem{b4_f} Retter A., Marom A., 2003, MNRAS, 345, L25

\bibitem{b4_g} Rich R.M., Mould J., Picard A., Frogel J.A., 
Davies R., 1989, ApJ, 341, L51

\bibitem{b4_h} Rushton M.T., Coulson I.M., Evans A., Nyman L.A., 
Smalley B., Geballe T.R., Van Loon J.Th., Eyres S.P.S., Tyne V.H., 
2003, A\&A, 412, 767

\bibitem{b4_h} Rushton M.T., Geballe T.R., Evans A., Smalley B.,
Van Loon J.Th., Eyres S.P.S., 2005a, MNRAS, 359, 624

\bibitem{b4_h} Rushton M.T. et al., 2005b, MNRAS, 360, 1281


\bibitem{b1} Sackmann I.J., Boothroyd A.I., Kraemer K.E., 1993, 
ApJ, 418, 457



\bibitem{b1} Schneider J., 2006, Extra-solar Planets Catalog,
http://cfa-www.harvard.edu/planets/cat1.html

\bibitem{b3_m} Shafter A.W., 1997, ApJ, 487, 226

\bibitem{b4} Shara M.M., Moffat A.F.J., 1982, ApJ, 258, L41

\bibitem{b4_i} Shara M.M., Moffat A.F.J., Webbink R.F., 1985, ApJ, 
294, 271

\bibitem{b4_j} Siess L., Livio M., 1999a, MNRAS, 304, 925

\bibitem{b4_k} Siess L., Livio M., 1999b, MNRAS, 308, 1133

\bibitem{b3_n} Siess L., Livio M., Lattanzio J., 2002, ApJ, 570, 329

\bibitem{b3_n} Siess L., 2006, A\&A, 448, 717




\bibitem{b4_o} Soker N., 1998, AJ, 116, 1308


\bibitem{b4_q} Soker N., Tylenda R., 2003, ApJ, 582, L105

\bibitem{b4_r} Tylenda R., 2004, A\&A, 414, 223

\bibitem{b4_r} Tylenda R., 2005, A\&A, 436, 1009

\bibitem{b4_r} Tylenda R., Soker N., 2006, A\&A, 451, 223

\bibitem{b4_r} Tylenda R., Crause L.A., G\'{o}rny S.K., Schmidt M.R.,
2005a, A\&A, 439, 651

\bibitem{b4_r} Tylenda R., Soker N., Szszerba R., 2005b, A\&A, 441, 1109

\bibitem{b4_s} Van Loon J.Th., Evans A., Rushton M.T., Smalley B.,
2004, A\&A, 427, 193 

\bibitem{b3_o} Wagner R.M., Starrfield S., 2002, IAUC, 7992

\bibitem{b3} Warner B., 1995, Cataclysmic Variable Stars, Cambridge 
University Press, Cambridge

\bibitem{b3} Wood P.R., Olivier E.A., Kawaler S.D., 2004, ApJ, 604, 800

\end{thebibliography}
\end{document}